\documentclass[jeosman,preprint,fleqn,showpacs,showkeys]{revtex4}
\usepackage{graphicx}
\usepackage{amssymb,amsmath,bm}

\begin{document}

\title{Hong-Ou-Mandel interference depends on the method of the eraseing the beam path information}

\author{ Sun-Hyun Youn  \footnote{E-mail: sunyoun@jnu.ac.kr, fax: +82-62-530-3369}}
\address{Department of Physics, Chonnam National University, Gwangju 500-757, Korea}

\begin{abstract}

  We study how the information of the beam path is related to the Hong-Ou-Mandel interference with two pulsed light  sources.  Through a simple model in which two photons in the form of pulses pass a beamsplitter and are observed at two detectors, we investigate how, during the measurement process, information about the paths of  the two photons can be erased. There are two ways to clear the information of the beam path, the first being  that from  the beginning, during the physical measurement process, the time information is not obtained.  The other is after measuring the information, to erase the temporal information in the  data analyzing process. We show that Hong-Ou-Mandel interference can  be obtained only when the beam path inofrmation is cleared from the physical measurement process.


\end{abstract}


\maketitle

\section{Introduction}

Hong-Ou-Mandel interference, which cannot be explained from a classical point of view, has become an important cornerstone for the development of quantum optics\cite{Hong-Ou-Mandel1987}.
Hong-Ou-Mandel interference is a subject that shows the pure quantum mechanical properties of light. Hong-Ou-Mandel interference was basically an experiment using two photons, but how the quantum mechanical properties of Hong-Ou-Mandel interference appear when more photons are included  is constantly being studied . Four-photon \cite{10Woolley}, and three-photon \cite{14Campos} conditions  are used to examine the Hong-Ou-Mandel interference. N identical single photon sources and measurement by N detectors in the far field system without beam splitter are also proposed to display the generalized N-photon Hong-Ou-Mandel interference\cite{15Campos}.

There is also experimentation with Hong-Ou-Mandel interference with quantum mechanical properties, even in systems that re composed of particles other than light. The realization of Hong-Ou-Mandel experiments using atoms instead of photons was reported \cite{7Lopes}. The two-phonon quantum interference experiment is also performed in a system of trapped-ion phonons\cite{8Toyoda},  two-particle interference in the electronic analog of the Hong-Ou-Mandel experiment in a quantum Hall conductor was investigated \cite{9Freulon},  and
general bosons and ferimions\cite{6Yuan} were considered to realize Hong-Ou-Mandel interference.
  
Relatively recently, experiments have been conducted on the interference of two photons whose modes do not  completely match.
Two-photon interference was performed with uncorrelated photons with different center frequencies from a luminescent light source \cite{17Kim}, and
two-photon interference of temporally seperated photons \cite{18Kim}. Also, a phase-randomized weak coherence state was used to perform generalized Hong-Ou-Mandel quantum interference\cite{20Zhang}.  A theoretical framework that describes Hong-Ou-Mandel interference for laser fields having arbitrary temporal waveforms and only partial overlap in time was developed\cite{3Agne}.

We study how the information of the beam path is related to the Hong-Ou-Mandel interference with two pulsed light sources. Through a simple model in which two photons in the form of pulses pass a beamsplitter and are observed at two detectors, we investigate how, during the measurement process, information about the paths of the two photons can be erased . In experiments, we usually get information from measurements and use them to analyze data. There are two ways to clear the information of the beam path; the first is that from the beginning, during the physical measurement process,  the time that  contains the path information is not obtained. The other way is to fter measuring the temporal information, to    erase the temporal information to remove the beam path information in the data analyzing process .

In a simple model, the time width of the pulse, adjusted to the frequency width of the filter placed in front of the detector, serves to erase information about the path of the pulse causing Hong-Ou-Mandel interference. On the other hand, if we delete the beam path information by ignoring the time information after measuring a short pulse without removing the pulse path information due to the time width increase by the filter, we cannot see the Hong-Ou-Mandel interference effect.
Through this, meaningful information in quantum mechanics is determined by the hamiltonian that interacts with light when measuring. Furthermore, after the measurement by the hamiltonian interacting has been completed, we  ignore or randomly  change the information, quantum phenomena such, as Hong-Ou-Mandel interference are not changed.

The present paper is organized as follows.
Section \ref{beamsplitter} explains the process of converting pulsed light in two different modes, which are already known, to two other lights through a beam splitter (Fig. \ref{FigSetup}). The time it takes for the two input lights to reach the beam splitter, and the time it takes for the light from the beam splitter to arrive at the detector, are  considered. The shape of the pulse in front of the detector is calculated by taking into account the pulse width of light, which is a Gaussian pulse form on the time axis, and the time width caused by the filter placed in front of the detector. The characteristics of the light in front of the detector are calculated for the general shape of the input light.

Section \ref{Hong-Ou-Mandel} calculates the probability that one photon will be detected by each detector in two different modes for single photon input in each mode.  
The probability of finding a photon at $\tau_c$   in the $ c$ mode detector and  $\tau_d$ in the $d$ mode detector is calculated considering the time width of the input pulse and the time width of the filter placed in front of the detector. By integrating over $\tau_c$ and $\tau_d$,  the probability of finding a single photon in each detector represents the well-known Hong-Ou-Mandel dip.
However, an interesting result can be obtained by taking the time integral over one of the two detectors, that is, $\tau_c$, and calculating the probability of a photon being found in the other detector as a function of time $\tau_d$. 
If the time width of the filter in front of one detector is significantly smaller than the time width of the input light and the two input light is large enough to decompose the two input light reaching the beam splitter, the Hong-Ou-Mandel dip is not visible regardless of the time width of the detector in front of the other detector. 
On the other hand, if the time width by the filter in front of one detector is slightly larger than the time width of the input light, the presence or absence of the Hong-Ou-Mandel dip is determined according to the time width by the filter in front of the other detector.
This is because when information on the path of a photon detected according to the time width of the filter placed in front of the detector is lost, the Hong-Ou-Mandel interference  is visible ;  and  when information on the path remains the  Hong-Ou-Mandel interference is not visible .
Importantly, the information on the path can be  erased according to  the size of the detector time window; but if there is no information on the path through the detector time window,  the Hong-Ou-Mandel interference does not occur.

Section \ref{conclusion} summarizes the main results, and discusses
their application. We discuss why the path erasure made by the time width change caused by the filter placed in front of the detector creates the Hong-Ou-Mandel interference, whereas path erasure by adjusting the size of the detector window does not make the Hong-Ou-Mandel interference.

\section{Traveling pulse through a Beam Splitter.} \label{beamsplitter}

\begin{figure}[htbp]
\centering
\includegraphics[width=10cm]{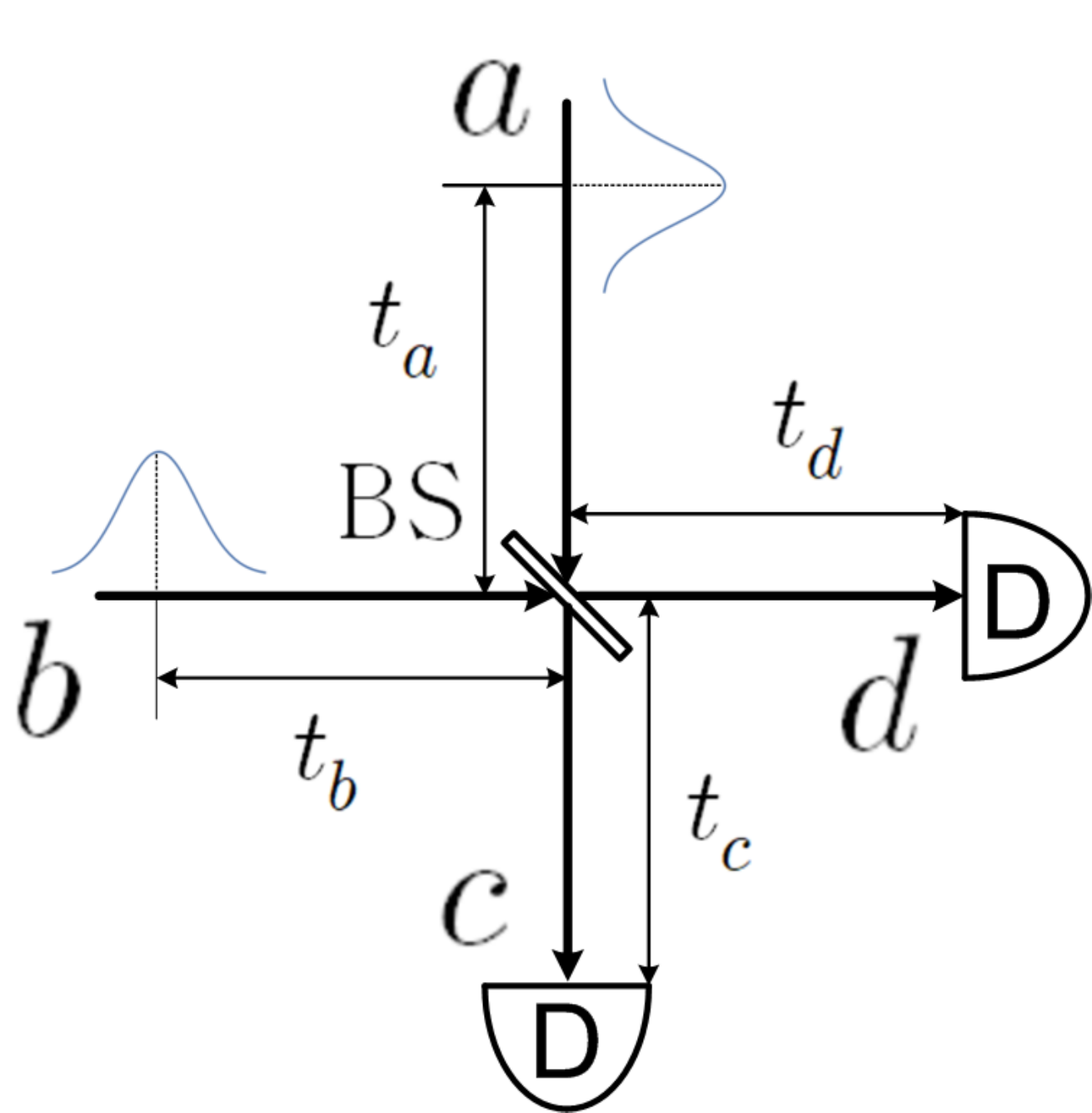}
\caption{After passing through the beam splitter (BS), two pulses in $a$ mode and $b$ mode are traveling into $c$  and $d$ modes .  It takes  $t_a $ ($t_b$)  time for a pulse in the $a$($b$) mode to go to  the beam splitter.  Pulses starting from the beam splitter take  $t_c$ ($t_d$) to go to the detector in $c$($d$) mode. D, detector. } \label{FigSetup}
\end{figure}

Let two modes $a$ and $b$ be coupled to $c$ and $d$ mode by a beam splitter, as in Fig. \ref{FigSetup}.  After passing through the beam splitter, two pulses in $a$ mode and $b$ mode are traveling into $c$  and $d$ modes .  It takes  $t_a $ ($t_b$)  time for a pulse in the $a$($b$) mode to go to  the beam splitter.  Pulses starting from the beam splitter takes  $t_c$ ($t_d$) to go to the detector in $c$($d$) mode. Using the time relation and the beam splitter relation \cite{Campos1989}.
\begin{eqnarray}
\hat  {a}(t')  &=&  t \hat{c} (t'- t_{ca}) -r  \hat{d} (t'-t_{da}),  \nonumber \\
\hat  {b}(t')  &=& r \hat{c} (t'- t_{cb}) + t  \hat{d} (t'-t_{db}),
\label{EqDefineAB}
\end{eqnarray}
where, $ t$ and  $r $ are the transmittance and the reflectance of the beam splitter, respectively. 
The time delay $t_{ij}$ is defind as  $t_{ij}  = t_{i} + t_{j} $ for  $(i,j) = a, b, c,d $. 

For simple caculation, we assume that the time behabier of the input beam is gaussian , and we define an operator that
has the same time-dependence as the field, i.e.  
\begin{eqnarray}
\hat  {a}(t')  &=&  \frac{\hat {a} }{\sqrt{\pi \delta_a ^2 } }e^{ -t'^2 / {\delta_a}^2  },  \nonumber \\
\hat  {b}(t')  &=&  \frac{\hat {b} }{\sqrt{\pi \delta_b ^2 } }e^{ -t'^2 / {\delta_b}^2  } .
\label{EqTimeAB}
\end{eqnarray}

With the Gaussian time behavier of two modes,  Eq. \ref{EqDefineAB} can be written 
\begin{eqnarray}
 \hat {a} e^{ -t'^2 / {\delta_a}^2  }   &=&  t  \hat {c} e^{ -(t' -t_{ca}) ^2 / {\delta_a}^2  }  
 -r   \hat {d}e^{ -(t' -t_{da}) ^2 / {\delta_a}^2  } ,  \nonumber \\
\hat {b} e^{ -t'^2 / {\delta_b}^2  }   &=&  r  \hat {c} e^{ -(t' -t_{cb}) ^2 / {\delta_b}^2  }  
+t   \hat {d} e^{ -(t' -t_{db}) ^2 / {\delta_b}^2  } , 
\label{EqDefineABtime}
\end{eqnarray}
where we assume  that  during traveling  to the detectors, the shape of the  input beam is not changed  .  

In general, an input state  can be written as:   
\begin{eqnarray}
 \psi_{in} &=&  \sum_{n,m}  c_n^a c_m^b |n>_a  |m>_b  \nonumber \\ 
&=& \sum_{n,m}   c_n^a c_m^b\frac{\hat{a^\dagger }^n \hat{b^\dagger}^m}{\sqrt{n! m!}}  |0>_a |0>_b
\label{EqINput}
\end{eqnarray}
where, we assumed that the input states at $a$ and $b$ modes  are $\psi_a =  \sum_n c^a_n |n> $ , $\psi_b =  \sum_m c^b_m |m> $, respectively.  If we use the time-dependent mode relation in  Eq. \ref{EqDefineAB},  Eq. \ref{EqDefineABtime} can be written:
\begin{eqnarray}
 \psi_{in} (t') = \sum_{n,m}  \frac{ c_n^a c_m^b } {\sqrt{n! m!} }& & \{ \frac{1}{\sqrt{\pi \delta_a ^2}}(t  \hat {c^\dagger } e^{ -(t' -t_{ca}) ^2 / {\delta_a}^2  }  
 -r   \hat {d^\dagger }e^{ -(t' -t_{da}) ^2 / {\delta_a}^2  }  )\} ^n  \nonumber  \\ & \times & 
\{ \frac{1}{\sqrt{\pi \delta_b ^2}}( r  \hat {c^\dagger } e^{ -(t' -t_{cb}) ^2 / {\delta_b}^2  }  +t   \hat {d^\dagger } e^{ -(t' -t_{db}) ^2 / {\delta_b}^2  } ) \}^m  |0>_c |0>_d
\label{EqINputNew}
\end{eqnarray}
 If  we put interference filters in front of the two detectors,  then  mode $c$ and $d$ should be changed.  In the time domain,  the interference filter with finite frequency width($\Delta \omega $) can be interpreted by the fourier transform limited time-dependent puse with finite  timewidth  ($\Delta t  \sim  \hbar  / \delta \omega $).  We can put:  
\begin{eqnarray}
F_i (t) =  \frac{1 }{\sqrt{\pi \delta_i ^2 } }e^{ -t^2 / {\delta_i}^2  }  ,
\label{EqTimeFilter}
\end{eqnarray}
where, $i = c, d $; and $\delta_i$ is the effective time width  introduced by the filter in the ($c, d$) modes. 
 
The input state in Eq. \ref{EqINputNew} can be written in terms of  the connvolution of interference filter  in Eq. \ref{EqTimeFilter}  as follows:
\begin{eqnarray}
  \psi_{in} (t') &=& \sum_{n,m}   \frac{ c_n^a c_m^b } {\sqrt{2n! m!} } \{  \hat {c^\dagger } \int_{-\infty}^{\infty} dt'' F_c (t'-t'') \frac{e^{ -(t'' -t_{ca}) ^2 / {\delta_a}^2  }}{\sqrt{\pi \delta_a ^2}}  
 -   \hat {d^\dagger }  \int_{-\infty}^{\infty} dt'' F_d (t'-t'')\frac{e^{ -(t'' -t_{da}) ^2 / {\delta_a}^2  } }{\sqrt{\pi \delta_a ^2}}   \} ^n  \nonumber  \\ & \times & 
\{   \hat {c^\dagger } \int_{-\infty}^{\infty} dt'' F_c (t'-t'') \frac{e^{ -(t'' -t_{cb}) ^2 / {\delta_b}^2  }}{\sqrt{\pi \delta_b ^2}}    +   \hat {d^\dagger }\int_{-\infty}^{\infty} dt'' F_d t'-t'') \frac{e^{ -(t'' -t_{db}) ^2 / {\delta_b}^2  }}{\sqrt{\pi \delta_b^2}}    \}^m  |0>_c |0>_d,
\label{EqINputCon}
\end{eqnarray}
where, in order to get a simple result,  we put $r = t = \frac{1}{\sqrt{2}}$ for the balanced  beam splitter . 
The convolution term can be calculated for the gaussian shape as:
\begin{eqnarray}
  G(t',i,j) &\equiv & \int_{-\infty}^{\infty} dt'' F_i (t'-t'') \frac{e^{ -(t'' -t_{ij}) ^2 / {\delta_j}^2  } }{\sqrt{\pi \delta_j^2}}    \nonumber \\
& = & \frac{1}{\sqrt{\pi(\delta_i ^2 + \delta_j ^2)}} e^{ - \frac{(t' - t_{ij})^2}{\delta_i ^2 + \delta_j ^2 }}.
\label{EqConvA2}
\end{eqnarray}

Then,  Eq. \ref{EqINputCon}  can be written as follows:
\begin{eqnarray}
  \psi_{in} (t') &=& \sum_{n,m}   \frac{ c_n^a c_m^b } {\sqrt{2n! m!} } \{  \hat {c^\dagger } G(t',c,a) 
 -   \hat {d^\dagger }  G(t',d,a)   \} ^n  \nonumber  \\ & \times & 
\{   \hat {c^\dagger } G(t',c,b) +   \hat {d^\dagger }  G(t',d,b)  \}^m  |0>_c |0>_d
\label{EqINputCon2}
\end{eqnarray}

In this section, we  obtained the output state $\psi_{in} (t')$  , traveling through a beam splitter, including the timde width of the input pulse and the filters.

\section{Time-Dependent Hong-Ou-Mandel Interference.} \label{Hong-Ou-Mandel}

For the pulse input states in $a$ and $b$ modes, we obtain 
the  output states in $c$ and $d$ modes  that  travel through  the beam splitter and filter. If we expand the powers in Eq. \ref{EqINputCon2} using the binomial coefficient, we obtain the following 
\begin{eqnarray}
 \psi_{out}   &=& \sum_{n = 0, m =0}  c_n ^a c_m ^b  \frac{1}{\sqrt{2 n! m!}} \sum_{p=0}^{n} \sum _{q=0}^{m}
   \frac{n!}{p !(n-p) !}\frac{m!}{q ! (m-q)!} \nonumber \\
   & \times &  G(t', c, a)^{p} (- G(t',d,a))^{n - p}  G(t',c,b)^{q}  G(t',d,b)^{m-q} \nonumber \\
 & \times& ({\hat c}^{\dagger})^{p+q}  ({\hat d}^{\dagger})^{n + m-p-q}  |0>_{c} |0>_{d}. \label{StOp3}
\end{eqnarray}
In order to measure the time-dependent probability, we  define the time $\tau_c$ and $\tau_d$ on two detectors in $c$ and $d$ modes, respectively.  
In order to obtain the probabilty that the detector measures a single photon in $c$  at $t'=\tau _c $ and $d$ mode at $t'=\tau_d$, we have to  find the coefficient such that $p+q=1$ and $n+m-p-q=1$ in Eq. \ref{StOp3}.  Only when $n+m=2$ can there  be a nonzero coefficient, the resilts being: 
\begin{eqnarray}
 \ <1(\tau_d)| <1 (\tau_c )| \psi_{out}> & =&  \sqrt{2} c_2 ^a c_0 ^b     G(\tau_c, c, a)   G(\tau_d,d,a)  + c_1 ^a c_1 ^b     G(\tau_c, c, b)   G(\tau_d,d,a)  \nonumber \\ &-& 
 c_1 ^a c_1 ^b     G(\tau_c, c, a)   G(\tau_d,d,b) - \sqrt{2}  c_0 ^a c_2 ^b     G(\tau_c, c, b)   G(\tau_d,d,b)   
, \label{StOp4}
\end{eqnarray}
where the time-dependence of  $G(t',i,j)$ is $G(\tau_c,i,j)$ if $i=c$  and  $G(\tau_d,i,j)$ if $i=d$, since the detection time is defined by the detector on each mode. 

If the the input state is two pulsed number state ($|1>_a |1>_b$) , then  Eq. \ref{StOp4} becomes;   
\begin{eqnarray}
 \ <1,1|1>_a |1>_b  & =&       G(\tau_c, c, b)   G(\tau_d,d,a) -  G(\tau_c, c, a)   G(\tau_d,d,b) .  \label{StOneOne1}
\end{eqnarray}
The joint probabilty that a detector  in the $c$ mode measures a single photon   and  a detector  in the $d$ mode measures asingle photon  becomes:
\begin{eqnarray}
 P_j (t_a, t_b, t_c, t_d , \tau_c , \tau_d )  &=&  | \ <1,1| \psi_{in}>|^2  \nonumber \\
 & =&  | G(\tau_c, c, b)   G(\tau_d ,d,a) -  G(\tau_c, c, a)   G(\tau_d,d,b) |^2 \nonumber \\
&=&  \frac{1}{2 \pi^2} |\{ \frac{e^{-\frac{(\tau_c-t_b -t_c )^2}{\delta_b^2+\delta_c^2} -\frac{(\tau_d-t_a -t_d )^2}{\delta_a^2+\delta_d^2}  }}{\sqrt{\delta_b^2+\delta_c^2} \sqrt{\delta_a^2+\delta_d^2}}-  \frac{e^{-\frac{(\tau_c-t_a -t_c )^2}{\delta_a^2+\delta_c^2} -\frac{(\tau_d-t_b -t_d )^2}{\delta_b^2+\delta_d^2}  }}{ \sqrt{\delta_a^2+\delta_c^2} \sqrt{\delta_b^2+\delta_d^2}} \}  |^2  . \label{StOneOne2}
\end{eqnarray}
\begin{figure}[htbp]
\centering
\includegraphics[width=10cm]{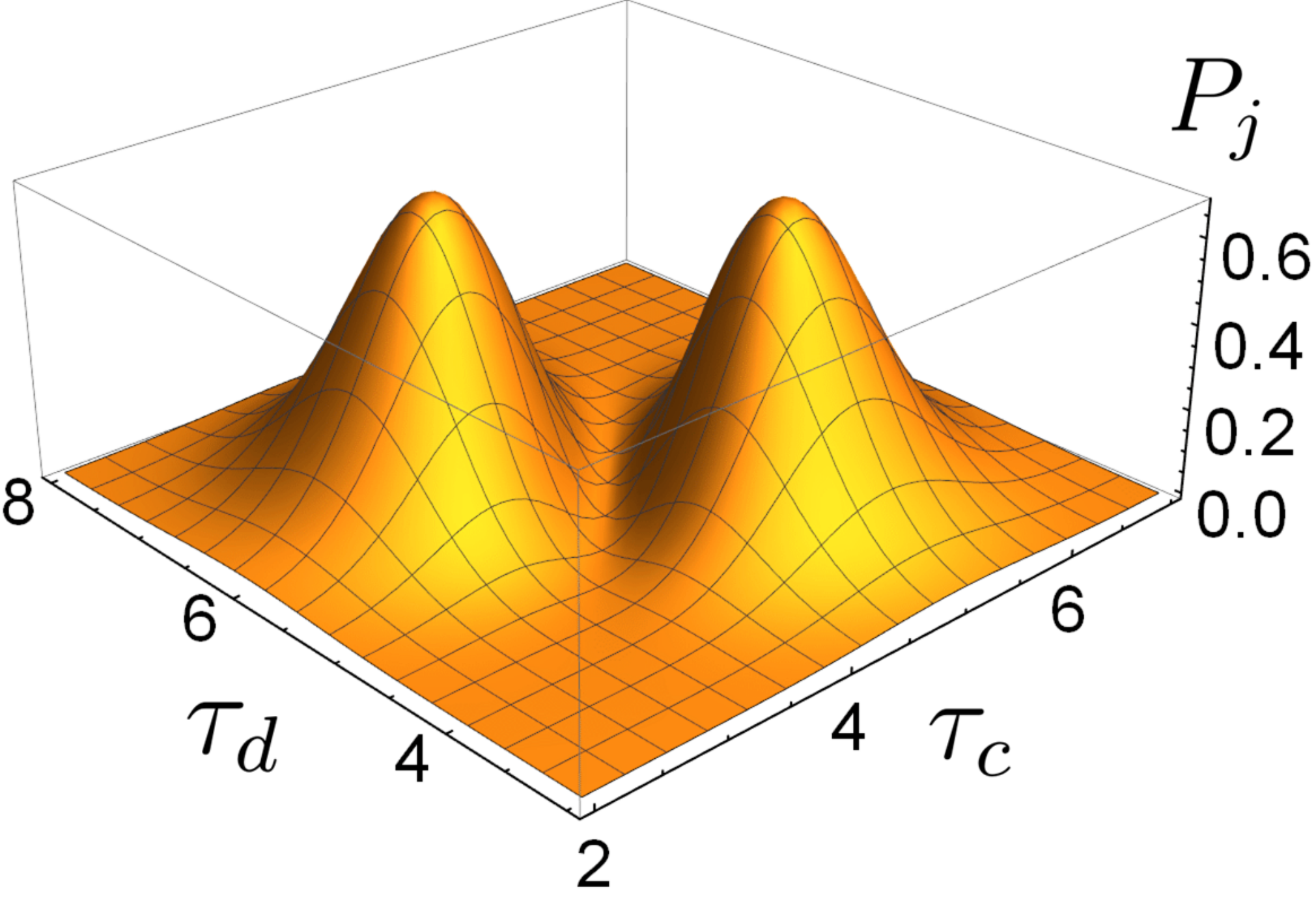}
\caption{Time-dependent joint probability ($P_j$) that each detector measures
a single photon at $\tau_c$ and $\tau_d$.  We set ($t_a, t_b, t_c, t_d $) as $(2,3,2,3)$ , where we use unit time scale. 
We also set all $ (\delta_a, \delta_b, \delta_c, \delta_d )$ as the same value $1$.  The actual scale of $P_j$ is $(\times 10^-2 )$.}
 \label{Fignotime1}
\end{figure}

  Fig. \ref{Fignotime1}  plots the joint probability $P_j$  that   each detector measures  a single photon at $\tau_c$ and $\tau_d$ in  Fig. \ref{Fignotime1}. 
For convinience,  we use unitless  time in this article;  in  other words,  time  $t$  in this articel represents  $t \times t_{unit} $, for example 
 $t_{uint} = 1  { ns}$.  The time delays ($t_a , t_b , t_c , t_d $) are set by $(2, 3, 2, 3)$ , the time widths  ($\delta_a , \delta_b $) of two input beams and
  the timewidth($\delta_c , \delta_b $)  related to the frequency filters in front of the  two detectors  are  all the same as $1$ in the Fig. \ref{Fignotime1} set-up.  In order to explain the time-dependence of $P_j$, we have to understand the interference effect in  Eq. \ref{StOp4}. The Interference effect can be estimated if we plot the pobability excluding the interference in Eq. \ref{StOneOne2}. Fig.   \ref{Fignotime1}  plots the probability   excluding the interference, i.e. we plot  $   |G(\tau_c, c, b)   G(\tau_d,d,a)|^2 + | G(\tau_c, c, a)   G(\tau_d,d,b)|^2 $. under the same parameters . The probabilty is the sum of two gaussian functions that  have peaks at  ($ \tau_c = 4 , \tau_d = 6$) and  ($ \tau_c = 5 , \tau_d = 5$).
    The peak at  ($ \tau_c = 4 , \tau_d = 6$) represents the case that a pulse from  $a$ mode is measured at the detector in $c$ mode and a pulse from $b$ mode is measured at the detector in $d$ mode.  If the two beams  change,  the probability will have  its  peak at    ($ \tau_c = 5 , \tau_d = 5$) accordingly.  
    If we reduce the time widths $(\delta s)$, the two peaks can be seperated and the probability  between two peaks  may be zero.  However, the prbability can be zero between the two peaks without changing the time widths.     
  If  the product of the two terms ( $  2 G(\tau_c, c, b)   G(\tau_d,d,a)   G(\tau_c, c, a)   G(\tau_d,d,b) $ is included,  we can see the dip in the middle of two gaussian peaks as in Fig. \ref{Fignotime1}.   This interference is basically related to the Hong-Ou-Mandel effect.

\begin{figure}[htbp]
\centering
\includegraphics[width=10cm]{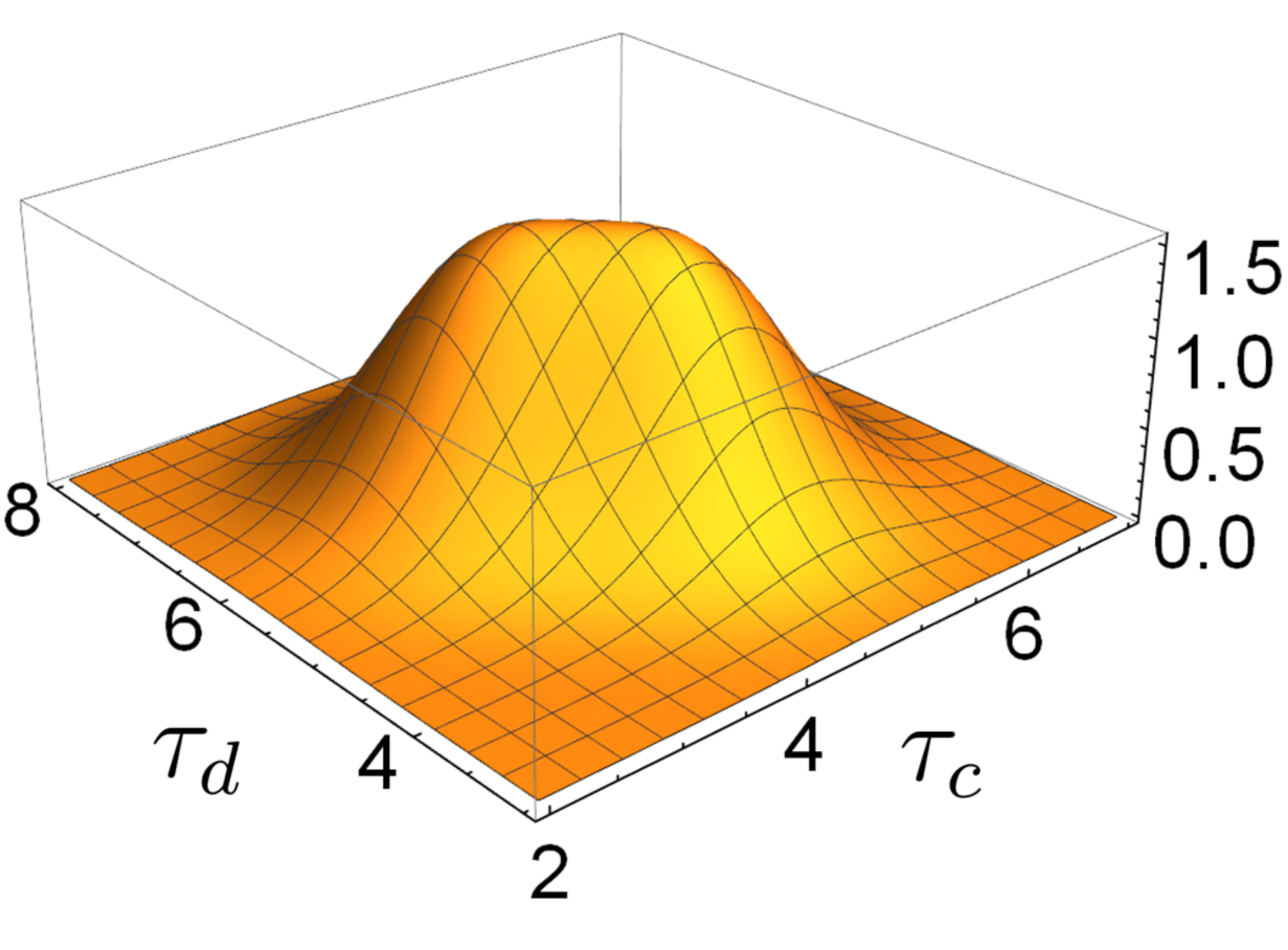}
\caption{Time-dependent joint probability  $( |G(\tau_c, c, b)   G(\tau_d,d,a)|^2 + | G(\tau_c, c, a)   G(\tau_d,d,b)|^2) $  that each detector measures
single photon at $\tau_c$ and $\tau_d$.  We set ($t_a, t_b, t_c, t_d $) as $(2,3,2,3)$ , where we use unit time scale. 
We also set all $ (\delta_a, \delta_b, \delta_c, \delta_d ) $ as the same value , $1$.  The actual scale of the probability is $(\times 10^-2 )$.}
 \label{FignotimeZero}
\end{figure}

In the usual Hong-Ou-Mandel interference  measurement, the probability can be calculated after integrating  the detector time $\tau_c$ and $\tau_d$, then 
\begin{eqnarray}
 P_j^{H} (t_a, t_b )  & = & \int_{-\infty} ^{\infty}  \int_{-\infty}^{\infty}   P_j (t_a, t_b, t_c, t_d , \tau_c , \tau_d )  d\tau_c d \tau_d  \nonumber \\
&=& \frac{1}{4\pi} \{ \frac{1}{\sqrt{ (\delta_b ^2 + \delta_c ^2 ) (\delta_a ^2 + \delta_d ^2)}}+\frac{1}{\sqrt{ (\delta_a ^2 + \delta_c ^2 ) (\delta_b ^2 + \delta_d ^2)}}  \nonumber \\ 
&-& \frac{4 e^{-\frac{2(t_a - t_b)^2 (\delta_a ^2 + \delta_b ^2 + \delta_c ^2 +\delta_d ^2 )}{(\delta_a ^2 + \delta_b ^2 +2 \delta_c ^2 )(\delta_a ^2 + \delta_b ^2 +2 \delta_d ^2)}}}
{\sqrt{(\delta_a ^2 + \delta_b ^2 +2 \delta_c ^2 )(\delta_a ^2 + \delta_b ^2 +2 \delta_d ^2)}}
\}.
 \label{EqHong-Ou-MandelPnotime}
\end{eqnarray}
 \begin{figure}[htbp]
\centering
\includegraphics[width=10cm]{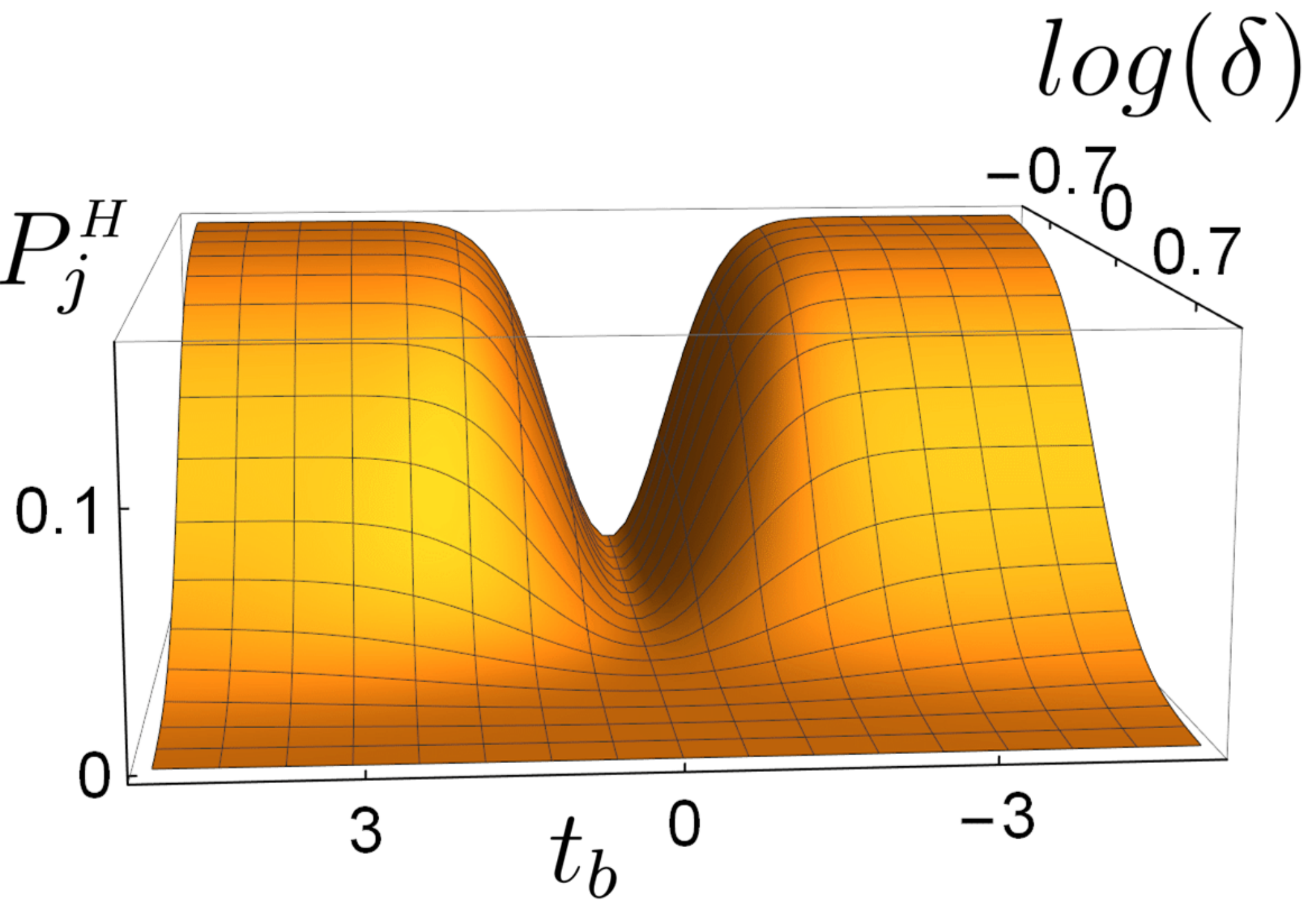}
\caption{Time joint probability ($P_j ^H$) that depends on the time delay $t_b$ and the timewidth  $\delta_c$ and $\delta_d$ . $\delta_c$ $(\delta_d)$ is related to the
 bandwidth of  filters in front of the detector inthe mode $c$ ($d$) by the fourire trasform relation.   ($t_a,  t_c, t_d $) are set as $(0,1,1)$ and    $ (\delta_a, \delta_b) $ are all set to $1$ and  $ (\delta_c ,  \delta_d )$ are all set to  $ \delta $.  } \label{FigDip1}
\end{figure}
Fig.  \ref {FigDip1} plots the  probability ($P_j ^H $) that depends on the time delay $t_b$ and the time width  $\delta_c$ and $\delta_d$ . 
 With fixing  ($t_a,  t_c, t_d $) as $(0,1,1)$, we change the time delay $t_b$. As $t_b$ approaches the value  $t_a = 0$, we can see the Hong-Ou-Mandel dip in the probability.  Since we integrate over the time $\tau_c$ and $\tau_d$, the probability $P_j^H$  has no dependence on $t_c$ and $t_d$.  
   In this figure, the time width of the  two input beams are set as $1$,  and we put the  timewidth related to the filters in $c$, $d$ modes as the same value $\delta$.
   The smaller the $\delta$, the deeper the  relative dip,  and the smaller  the width of the dip.   It is clear that as the time width  $\delta$ increases, the width of the dip increases.   On the other hand, if the time width $\delta$ narrows, the question might arise of  whether the interference effect will be erased or not. In fact, the interference effect is not erased,  because the time width(1) of the two input pulses already has sufficient interference effect at $ t_b-t_a = 2 $.

An ineresting result can be obtaind if we integrate only over $\tau_d$ for the joint probabilty $P_j^D (t_a, t_b, t_c, t_d , \tau_c , \tau_d ) $, then:
\begin{eqnarray}
 P_j ^D (t_a , t_b ,t_c, t_d, \tau_c)  & = &  \int_{-\infty}^{\infty}   P_j (t_a , t_b , t_c , t_d, \tau_c , \tau_d ) d \tau_d  \nonumber \\
&=& \frac{1}{4 \pi^{3/2}} \{ \frac{\sqrt{2} e^{-\frac{2(\tau_c - t_b - t_c)^2 }{\delta_b ^2 + \delta_c ^2}}}{ (\delta_b ^2+ \delta_c ^2)\sqrt{\delta_a ^2 +\delta_d ^2 }}+ \frac{\sqrt{2} e^{-\frac{2(\tau_c - t_a - t_c)^2 }{\delta_a ^2 + \delta_c ^2}}}{ (\delta_a ^2+ \delta_c ^2)\sqrt{\delta_b ^2 +\delta_d ^2 }} \nonumber \\
&-&  \frac{4 }{\sqrt{(\delta_a ^2 +  \delta_c ^2)(\delta_b ^2 +
\delta_c ^2)(\delta_a ^2 + \delta_b ^2 + 2 \delta_d ^2)}}  e^{-H(t_a, t_b, t_c, t_d,  \delta_a, \delta_b, \delta_c, \delta_d )}
\} ,  \label{EqHong-Ou-MandelPtauC}
\end{eqnarray}
where, $ H $ is defined as:
\begin{eqnarray}
  H (t_a, t_b, t_c, t_d,  \delta_a, \delta_b, \delta_c, \delta_d ) &=&  \frac{(\tau_c - t_a -t_c)^2}{\delta_a ^2 + \delta_c ^2}+
\frac{(\tau_c - t_b -t_c)^2}{\delta_b ^2 + \delta_c ^2} +\frac{(t_a +t_d)^2}{\delta_a ^2 + \delta_d ^2}+\frac{(t_b +t_d)^2}{\delta_b ^2 + \delta_d ^2} \nonumber \\
& -& \frac{(t_b (\delta_a ^2+ \delta_d ^2 ) +  t_a (\delta_b ^2 + \delta_d ^2 ) + t_d (\delta_a ^2 + \delta_b ^2 + 2 \delta_d ^2)^2)}
{ (\delta_a ^2+ \delta_d ^2 ) (\delta_b ^2 + \delta_d ^2 ) (\delta_a ^2 + \delta_b ^2 + 2 \delta_d ^2)} . \label{EqHftn}
\end{eqnarray}

 \begin{figure}[htbp]
\centering
\includegraphics[width=10cm]{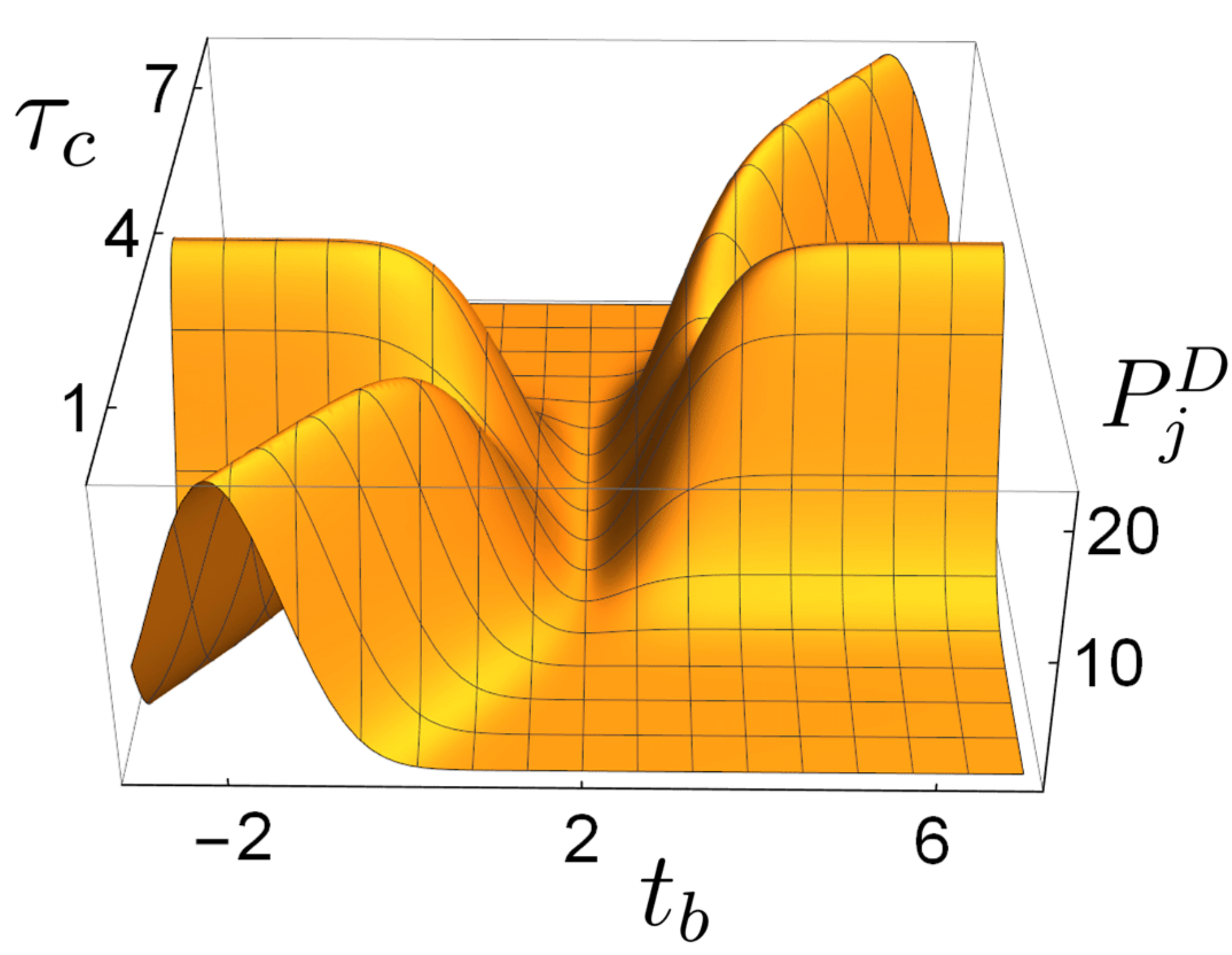}
\caption{Time-dependent  joint probability ($P_j ^D $) that a detector in $c$ mode  measures
a single photon at $\tau_c$ as a function of  $t_b$.  To obtain $ P_j ^D $, we integrate over the delay time $\tau_d$.   
We set ($t_a, t_c, t_d $) as $(2,2,3)$ , where we use unit time scale. 
We set all $ (\delta_a, \delta_b, \delta_c, \delta_d )$  to the same value, $1$.  The actual scale of $P_j ^D$ is $(\times 10^{-3} )$.} \label{FigtBTauC1}
\end{figure}

 Figure  \ref{FigtBTauC1} plots the time-dependent  joint probabilities ($P_j ^D $) where the detector in $c$ mode measures a single photon at  $\tau_c $  and  delay time  $t_b$.  All  time widths are as set to the same value $1$, and  after integration, the time-dependence $\tau_c $ has been removed .  The dip in the central region ($t_b = 2 $) is directly connected to the Hong-Ou-Mandel dip. As $t_b$ moves away from $t_b = 2$, as $t_c$ changes,  we can see two peaks.
For example,  for $t_b = 5$, there are two peaks  at  $\tau_c  = 4 $ and $\tau_ c  = 7$. The peak of $\tau_c  = 4$ occurs when a pulse in $a$ mode is measured in $c$ mode and a pulse in $b$ mode is measured in $d$ mode.  The other peak at   $\tau_c  = 4 $ is related to the case where the pulse from $a$ mode is measured in $d$ mode.

  \begin{figure}[htbp]
\centering
\includegraphics[width=10cm]{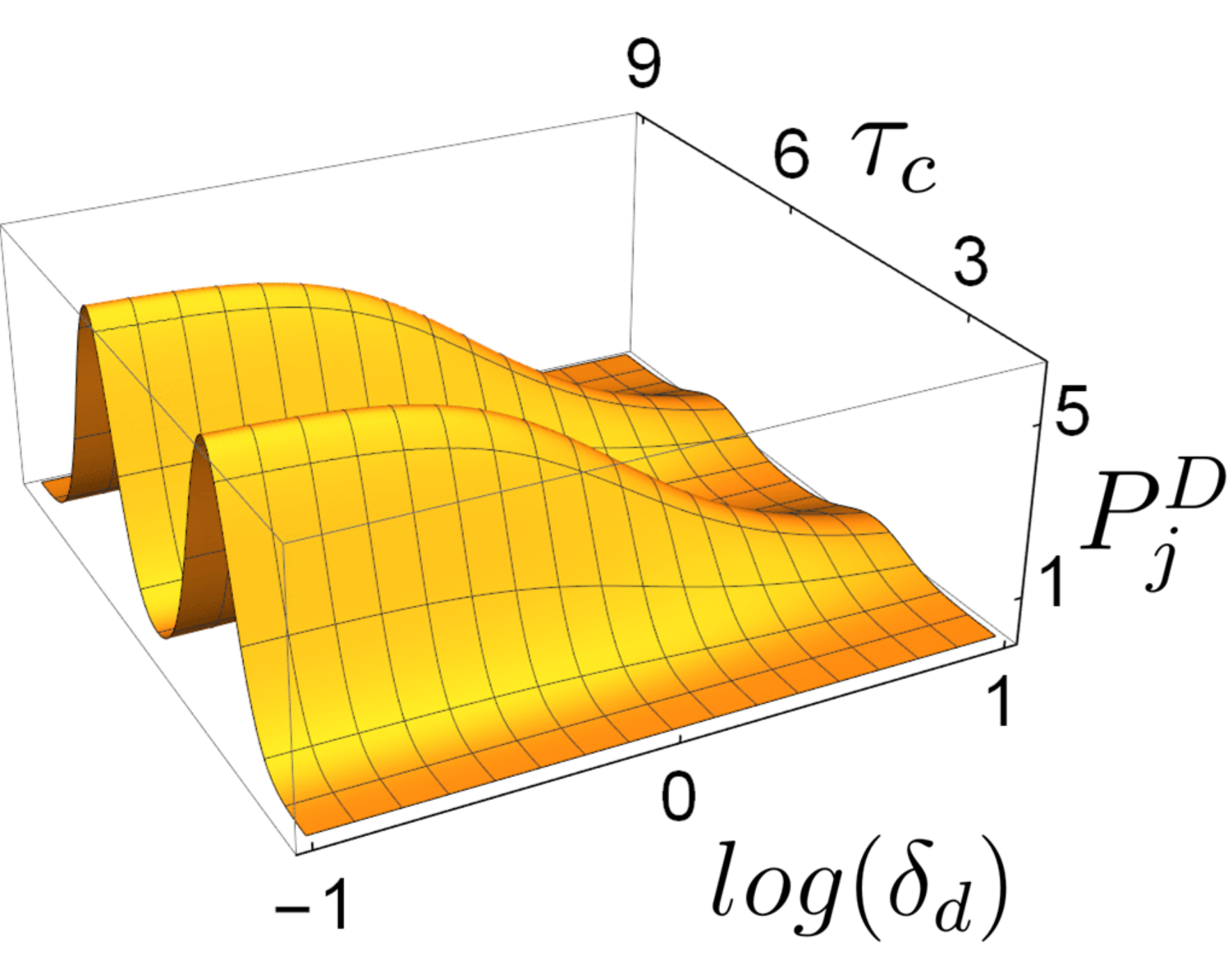}
\caption{Time-dependent joint probability ($P_j ^D$) that a detector in $c$ mode  measures a
single photon at $\tau_c$ as a function of the timewidth  $\delta_d$.   The time width $ \delta_c $ is $0.1$.
To obtain $ P_j$, we integrate over the delay time $\tau_d$.   
We set ($t_a, t_b, t_c, t_d $) as $(2,5,2,2)$ , where we use unit time scale. 
We also set all $ (\delta_a, \delta_b) $ as the same value $(1, 1)$. 
The actual scale of $P_j ^D$ is $(\times 10^{-2} )$.} \label{FigTauCA1}
\end{figure}

Figure  \ref{FigTauCA1} plots the probability $P_j ^D$ as a function of $\tau_c$ and time width $\delta_d$.  The time widths of the two input pulses are set to the same value $1$,  and we also set the time width  $\delta_c$ as $0.1$.  For the time width $\delta_d  = 0.1$ , the time delay $3$ between two input pulses as $\tau_c$ changes gives two peaks. This is because the time width is short enough to seperate the two peaks.  As $\delta_d $ increases, the overlapping part of the two pulses widens.   However, the separation between the two peaks continues, even though the time width $\delta_d$ is increased by $10$.
 If we measure photons on the detector with a time width of $10$ in $d$ mode, we can not tell if the photon is from $a$ mode or $b$ mode. However, for the detector in $c$ mode  whose time width is $0.1$, we  can check the origin of the measured photon. Therefore, in this setup, we cannot see the Hong-Ou-Mandel interference effect.   The dip between the two peaks comes from the fact that these two peaks are separate independent  Gaussian peaks.

  \begin{figure}[htbp]
\centering
\includegraphics[width=10cm]{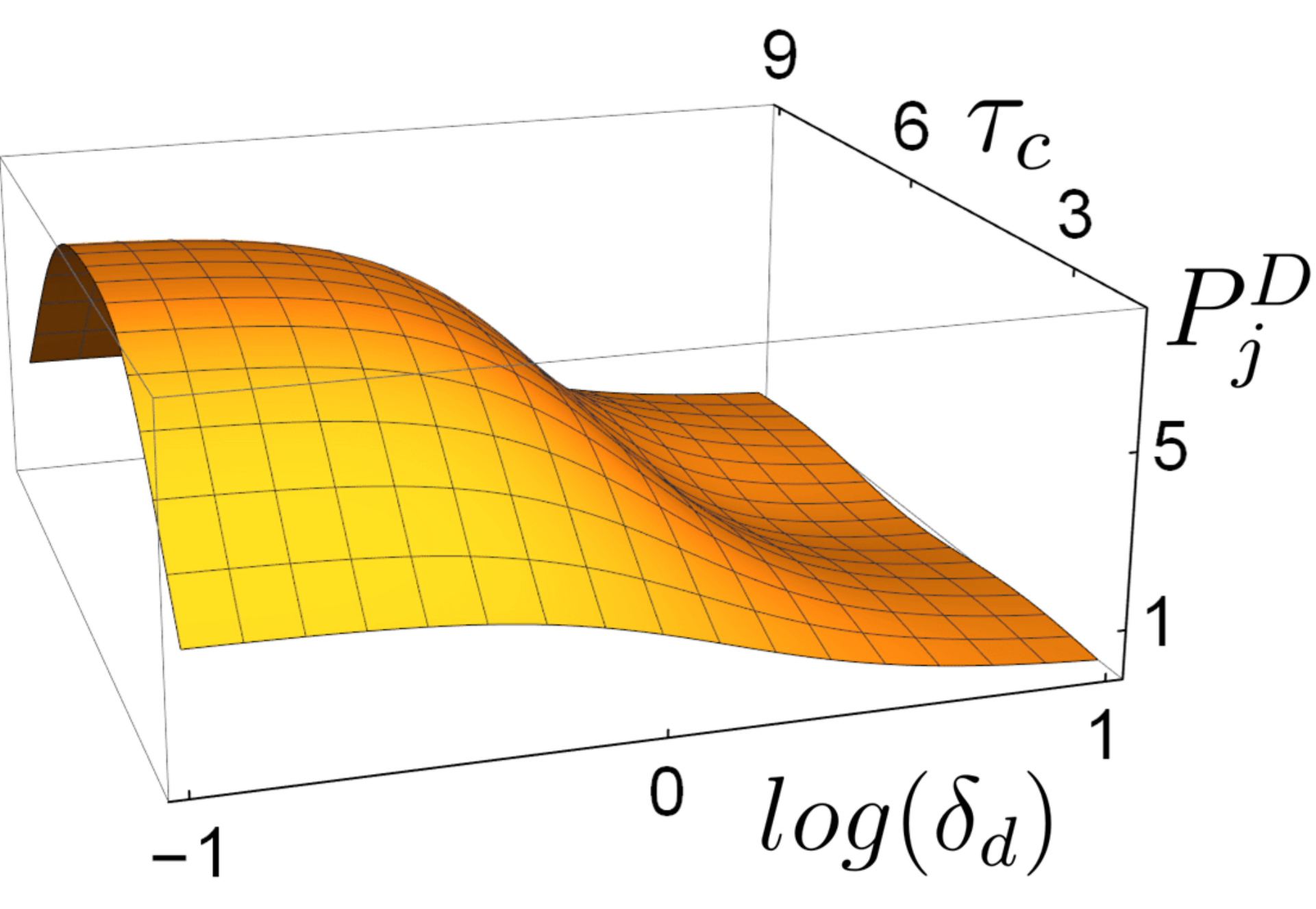}
\caption{Time-dependent joint probability ($P_j^ D$) that a detector in $c$ mode  measures
a single photon at $\tau_c$ as a function of the timewidth  $\delta_d$.   The time width $ \delta_c $ is $3$.
 To obtain $ P_j ^D$, we integrate over the delay time $\tau_d$.   
We set ($t_a, t_b, t_c, t_d $) as $(2,5,2,2)$ , where we use unit time scale. 
We also set all $ (\delta_a, \delta_b, \delta_c ) $ as the same value, $(1, 1, 3)$. 
The actual scale of $P_j ^D$ is $(\times 10^{-3} )$.} \label{FigTauCB1}
\end{figure}

\begin{figure}[htbp]
\centering
\includegraphics[width=10cm]{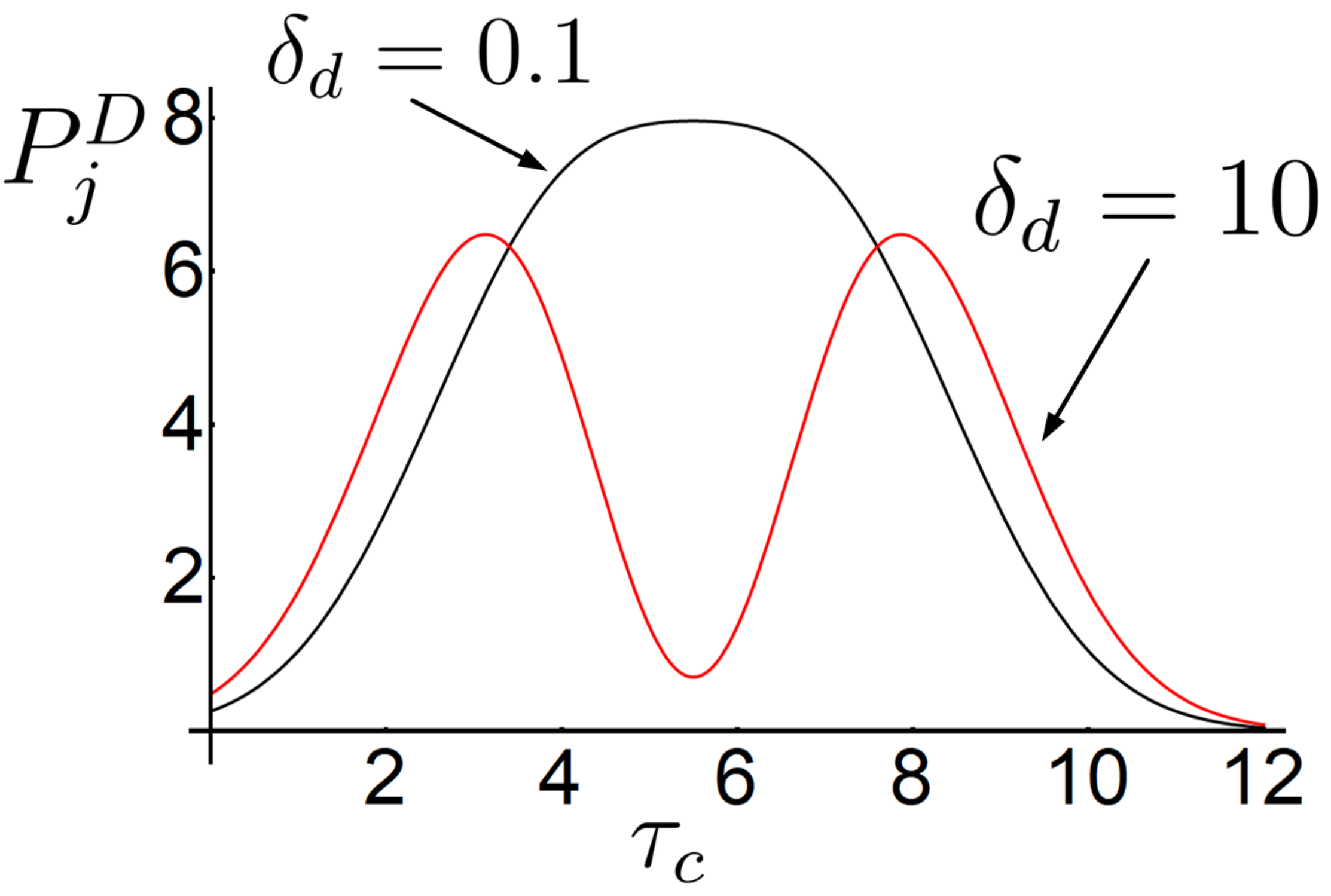}
\caption{Time-dependent joint probability ($P_j ^D$) that a detector in $c$ mode  measures
a single photon at $\tau_c$ . We plotted two set of data for the time width  $\delta_d=0.1 $ and    $\delta_d=10 $. 
 To obtain $ P_j ^D$, we integrate over the delay time $\tau_d$.   
We set ($t_a, t_b, t_c, t_d $) as $(2,5,2,2)$ , where we used unit time scale. 
We also set all $ (\delta_a, \delta_b, \delta_c $ as the same value $(1, 1, 3)$. 
The actual scale of $P_j ^D$ is $(\times 10^{-3} )$ for $\delta_d=0.1 $  and  $(\times 5 \times  10^{-4} )$ for $\delta_d=10 $ . } \label{FigTauCAB1}
\end{figure}

 Figure \ref{FigTauCB1} plots the probability $P_j ^D$ as a function of $\tau_c$ and time width $\delta_d$.  The time widths of two the input pulses are set to the same value $1$,  and  the time width  $\delta_c $  is also set to $3$.  The time delay $3$ between the two input pulses might  give two peaks as $\tau_c$ changes, but the time width $\delta_d = 3$ is not short enough to make a seperation of two peaks with time distance  ($t_b  -  t_a = 3$).  Although  the two probability peaks seem to be close enough to make a  Hong-Ou-Mandle dip, there is no dip between the two peaks.  This is because  the time width $\delta_d =0.1$ is sufficient  to give the path of the photon measured   at the detector in $d$ mode. Although the increase of  $\delta_d $  results in a greater overlapping part of the two pulses,  as $\delta_d $ increases, a dip is evident in the two peaks. In order to clearly see the dip in the two peaks, Fig \ref{FigTauCAB1} plots the joint probability ($P_j ^D$) that a detector in $c$ mode  measures a single photon at $\tau_c$   for the time widths  $\delta_d =0.1 $ and    $\delta_d =10 $.  
 The Hong-Ou-Mandel dip is associated with the interference of two photons only if the path of the two photons cannot be distinguished. 
 The path of the photon measured at the detector in $c$ mode cannot be distinguished since the time width $(\delta_c = 3 )$ is not short enough to seperate the two paths; furthermore,  the time width ($\delta_d = 10$) is long enough to erase the path information. Therefore, the dip between two peakses for   $\delta_d =10 $
came from the quantum interferences.

   The pulse width of the two photons is $\delta_a  =  \delta_b  = 1$, and the time distance between the two pulses is $(t_b - t_a) = 3$, so there is no chance for the two photons to meet together in the time domain. Where does the interference come from? We might say that  the path information is  important. Although the two pulses are well seperated in the time domain, some uncertainty was introduced by the interference filter whose frequency bandwidth  was related to the time width.  The convolution effect  erases the path information. 
   
   We now change the setup slightly. We want to add a time window for a detector in mode $d$ for the new setup.   If we add all the data for that window, we might delete  the arrival time of the photon; then, the Hong-Ou-Mandel interference  may be produced for certain conditions. 
   In other words, for the time width $\delta=0.1$ in Fig. \ref{FigTauCA1}, if we open the window much wider than the time width such as $10$, we cannot obtain the arrival time of the photon.  Then the path information is deleted. This might make  quantum interference.
   
 We calculate the joint probability that the time window for the detector in the $d$ mode is $(\tau_d - t_w , \tau_d + t_w$) as follows:

\begin{eqnarray}
 P_j ^W  &( &t_{ab} , \tau_c, \tau_d, t_w)  =  \int_{\tau_d - t_w}^{ \tau_d + t_w }   P_j (t_{ab} ,0,0,0, \tau_c , {\tau_d}'  ) d {\tau_d}'  \nonumber \\
&=& \frac{1} {8 \sqrt{2} \pi^{3/2} (1+\delta_c ^2) \sqrt{1+ \delta_d ^2}} e^{- \frac{2 (t_{ab} ^2 - \tau_c ^2 (1+\delta_d ^2))}{(1+\delta_c ^2)(1+\delta_d ^2 )}} 
[  e^{ \frac{2 t_{ab} ^2}{ (1+\delta_c ^2)(1+\delta_d ^2 )}} G(t_w , t_{ab} - \tau_d , 1+ \delta_d ^2 ) \nonumber \\ 
&+&e^{ \frac{2 t_{ab} (-t_{ab} \delta_d ^2 +2 \tau_c (1+ \delta_d ^2))}{ (1+\delta_c ^2)(1+\delta_d ^2 )}}  G(t_w, \tau_d , 1+ \delta_d^2 ) \nonumber \\ &-&
 e^{ \frac{t_{ab} \{4 \tau_c (1+ \delta_d ^2 ) - t_{ab} (-1+ \delta_c ^2 + 2 \delta_d ^2)\}}{2 (1+\delta_c ^2)(1+\delta_d ^2 )}} G(2 t_w , t_{ab}- 2 \tau_d , 4(1+ \delta_d ^2 ))
]
 \label{Eq2Win}
\end{eqnarray}
     where,
\begin{eqnarray}
 G(t_1 ,  t_2, \Delta ) =  \frac{(t_2 + t_1 )}{\sqrt{(t_2 + t_1 )^2}} Erf(\sqrt{ \frac{2 (t_1 + t_2 )^2}{ \Delta}})-   \frac{(t_2 - t_1 )}{\sqrt{(t_2 - t_1 )^2}} Erf(\sqrt{ \frac{2 (t_2 -  t_1 )^2}{ \Delta}})
 \label{Eq2Gftn}
\end{eqnarray}
    and the function $Erf(x)$ is defined  by the value $ \frac{2}{\sqrt{\pi}} \int _{0}^{x} e^{-t^2} dt $. We also set   $t_i = 0 $ for all the time delays, except $t_a  = t_{ab}$ , as  for simplicity, we also assume that the initial input pulse has time width $ \delta_a =  \delta_b  = 1$.
     
\begin{figure}[htbp]
\centering
\includegraphics[width=10cm]{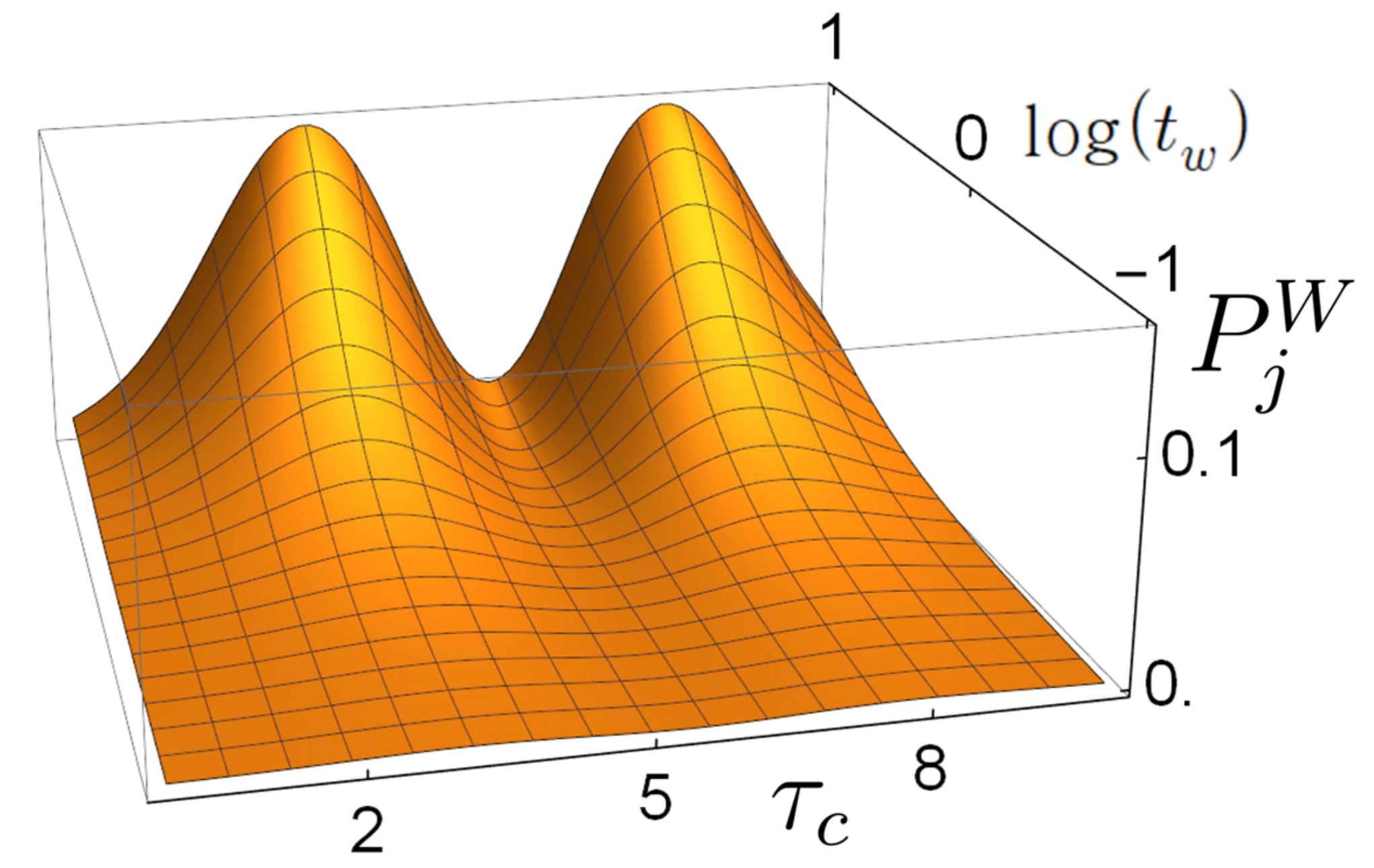}
\caption{Time dependent joint probability ($P_j ^W$) that a detector in $c$ mode  measures
single photon at $\tau_c$  with  the time window  ($5.5 -t_w , 5.5+t_w $).  The time width related to the filter in $c$( $d$)  mode is   $\delta_c =3 $ 
($\delta_d =10 $). 
 We set ($t_a, t_b, t_c, t_d $) as $(2,5,2,2)$  and we also set   $ (\delta_a, \delta_b)$ as to the same value $1$. 
The actual scale of $P_j ^W$ is $(\times 10^{-3} )$  } \label{FigWindowL}
\end{figure}

      Figure \ref{FigWindowL} plots the joint probability for the time window  ($\frac{t_{ab}}{2} -t_w , \frac{t_{ab}}{2}  + t_w $).  The time widths $\delta_c $ and $\delta_d$ are ($3$ and $10$), respectively.   Figure  \ref {FigTauCAB1} shows  the same condition for the graph with $\delta_d  = 10 $.  The same result can be obtained if the time window $t_w$ increases by $10$.  If the time window $t_w$ is small, the joint probability is small since we integrate all data within the time window  ($\frac{t_{ab}}{2} -t_w , \frac{t_{ab}}{2} + t_w $).  The dip in the middle of two peaks is comes from the quantum interference  for the same reason explained in Fig. \ref{FigTauCB1}.  Actually, the path information is already lost, since   regardless of the time window, the time width $\delta_d $ is $10$. 
          
\begin{figure}[htbp]
\centering
\includegraphics[width=10cm]{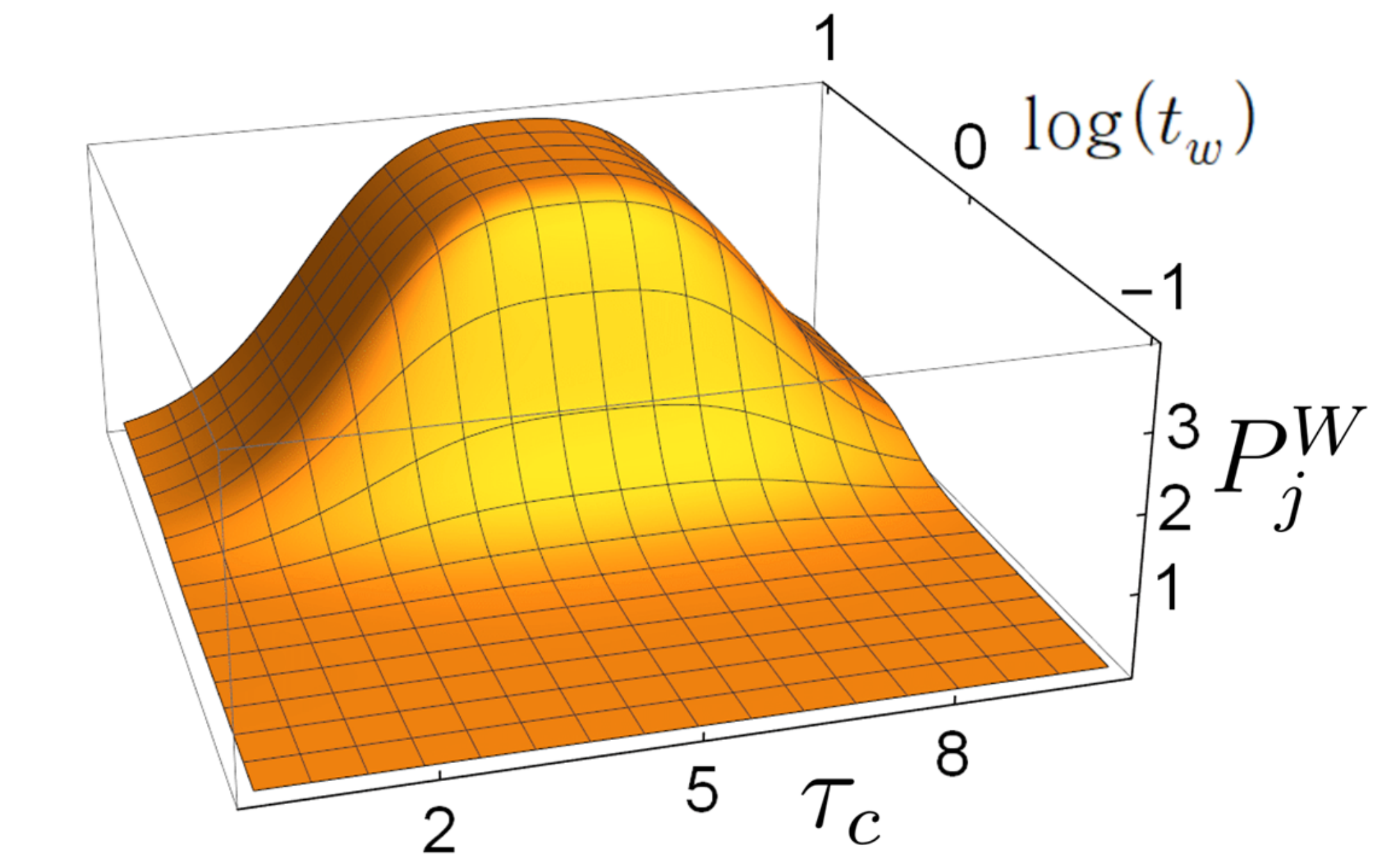}
\caption{Time-dependent joint probability ($P_j ^W$) that a detector in $c$ mode  measures
a single photon at $\tau_c$  with  the time window  ($5.5 -t_w , 5.5+t_w $).  The time width related to the filter in $c$( $d$)  mode is   $\delta_c =3$ 
 $ (\delta_d =0.1) $. 
 We set ($t_a, t_b, t_c, t_d $) as $(2,5,2,2)$  and we also set   $ (\delta_a, \delta_b)$  to the same value, $1$. 
The actual scale of $P_j ^W$ is $(\times 10^{-3} )$  } \label{FigWindowS}
\end{figure}

 Figure \ref{FigWindowS}  plots the joint probability for the time window  ($\frac{t_{ab}}{2} -t_w , \frac{t_{ab}}{2} + t_w $).  The time width $\delta_c $ and $\delta_d$ are ($3$ and $0.1$), respectively. Eventhough, the time width $\delta_d$ is $0.1$, when we integrate all the data from the window ($\frac{t_{ab}}{2} -t_w , \frac{t_{ab}}{2}  + t_w $), we cannot obtain the path information of the two input pulsesIf $t_w$ is $10$, then the information from the detector in $d$ mode is simply the fact that a photon was measured.  There is no information as to whether the measured photon came from $a$ mode or $b$ mode. However, the calculation shows that even if the path information of the measured photons is not known, there is no quantum interference.  The result is very similar to the data in Fig. \ref {FigTauCAB1} for  the graph with $\delta_d  = 0.1 $ 

At first glance, we think opening the time window up to $t_w = 10$ can lead to quantum interference, because we do not know the path information of the detector's photons in $d$ mode. From a photon's standpoint, it is not important whether the exact time that the photon was measured is recorded or not. The important thing is that the photon was measured at the detector in $d$ mode after passing the filter with time width $\delta_d = 0.1$.

\section{Conclusion and Discussion.} \label{conclusion}

 We investigated how information of the  paths of two photons can be erased during the measurement process, with a simple model in which two pulse-shaped photons pass a beamsplitter, and are measured at two detectors, 

First, we calculate the time-dependent joint probability  $P_j $  that each detector measures a single photon at a different time. The joint probability is not a simple sum of Guassian probability, as there is the interference  effect causd by the uncertainty of the beam path of the two pulses. If we integrate the time-dependent joint probability over the detection time in each detector, we can obtain the normal Hong-Ou-Mandel interferece effect. 

We calculated the probability  $P_j ^D$  obtained  by integrating $P_j$ over the detecting time $\tau_d $  in the $d$ mode as a function of $\tau_c$ and time width $\delta_d$.  The time widths of the two input pulses are set  to the same value $1$ and the time distance   ($t_b - t_a = 3$).  When the time width $ \delta_c  = 0.1$ as in Fig. \ref{FigTauCA1},  the time delay $3$ between two input pulses gives two peaks as $\tau_c$ changes for the time width $\delta_d  = 0.1$. This is because the time width is short enough to seperate the two peaks.  As $\delta_d $ increase, the overlapping part of two pulses widens.   However, the separation between the two peaks continues, even though the time width $\delta_d$ increases by $10$. If we measure photons on the detector with a time width of $10$ in $d$ mode, we cannot tell if it is from $a$ mode or $b$ mode. However, for the detector in $c$ mode  whose time width is $0.1$ can check the origin of the measured photon. Therefore, in this setup, we cannot see the Hon-Ou-Mandel interference effect.   The dip between the two peaks came from the fact that these two peaks are separate independent Gaussian peaks.  

On the other hand,  when the time width $ \delta_c  =3 $,  as in Fig. \ref{FigTauCB1}, the time delay $3$ between two input pulses might  give two peaks as $\tau_c$ changes, but the time width $\delta_c = 3$ is not short enough to make a seperation of two peaks with time distance  ($t_b - t_a = 3$).  Although the two probability peaks seem to be close enough to make a  Hong-Ou-Mandle dip, there is no dip between the two peaks for the time width $\delta_d = 0.1$.  This is  because the time width $\delta_d =0.1$ is sufficient  to give us the path of the photon measured  at the detector in $d$ mode.   Although the increase of  $\delta_d $  increases the overlapping part of the  two pulses in the middle of the  two peaks,  we can see a clean dip between the two peaks as   $\delta_d $  increases. In  Fig \ref{FigTauCAB1},  we can see the dip between the two peaks as $\tau_c $ changes.  The Hong-Ou-Mandel dip is associated with the interference of two photons only if the path of the two photons cannot be distinguished.  The path of photon measured at the detector in $c$ mode can not be distinguished since the time width $(\delta_c = 3 )$ is not short enough to separate the two pathes; furthermore, the time width ($\delta_d = 10$) is long enough to erase the path information. Therefore, the dip between the two peaks for      $\delta_d =10 $  is comes from the quantum interferences. 

Although both the pulse width of the two photon is $\delta_a  =  \delta_b  = 1$ and the time distance between the two pulses is $(t_b - t_a) = 3$, this  gives no chance for the two photons meet together in the time domain, as some uncertainty was introduced by the interference filter whose frequency bandwidth  was related to the time width. The lost path information  gives Honfg-Ou-Mandel interference.

   We changed the setup and  added a time window for a detector in mode $d$.  For the time width $\delta_c  = 3,  \delta_d =0.1 $ in Fig. \ref{FigWindowS}, we increased the width of the time window up to $t_w =10$;  we then cannot obtain the arrival time of the measured photon in $d$ mode, in other words, we have no beam path information of the measured photon in $d$ mode.  However, we can not see the Hong-Ou-Mandel interference.  For the same time window $t_w =10 $,  the Hong-Ou-Mandel interference can be seen for the time width   $\delta_c  =3,   \delta_d  =10 $  as in Fig. \ref{FigWindowL}.   The change in the size of the window only affects the change in the size of the signal and does not change the shape of the signal.  
        
If the time width of the pulse modified by  the filter placed in front of the detector is large enough to erase the beam path information,  Hong-Ou-Mandel interference can be made although the pulse width  of the initail pulse is short enough to separate the two pulses at the detector.   On the other hand, if we delete the beam path information by  ignoring the time information after measuring  short pulses  we  cannot see the Hong-Ou-Mandel interference effect.

Through this, meaningful information in quantum mechanics is determined by the Hamiltonian that interacts with light when measuring. Furthermore, 
even if we randomly ignore or change the information after the measurement by the Hamiltonian interaction has been completed, quantum phenomena, such as the Hong-Ou-Mandel interference are not changed. In our study, we considered a situation where none of the characteristics of the detector were changed, depending on the wavelength of light


\begin{references}

\bibitem{Hong-Ou-Mandel1987} C. K. Hong, Z. Y. Ou, and L. Mandel  Phys. Rev. Let. {\bf 59}, 2046  (1987)
T. UMAP. Journal {\bf 25.4}, 378 (2004).




\bibitem{10Woolley}  M. J. Woolley, C. Lang, C. Eichler, A. Wallraff and A Blais,New. Jo. of Phys.  {\bf 15} 1367 (2013)

\bibitem{14Campos} R. A. Campos, Phys. Rev. A,  {\bf 62} 1050 (2000)

\bibitem{15Campos} S. Mahrlein, S. Oppel, R. Wiegner and  J. von Zanthier J. of Mod. Opts.  {\bf 64} 921 (2017)




\bibitem{7Lopes} R. Lopes, A. Imanaliev, A. Aspect, M. Cheneau, D. Boiron and C. C. Westbrook, Nature,  {\bf 520}, 66 (2015)

\bibitem{8Toyoda} K. Toyoda, R. Hiji, A. Noguch and S. Urabe,  Nature,  {\bf 527}, 74 (2015)

\bibitem{9Freulon} V. Freulon, A. Marguerite, J. M. Berroir, B. Placais, A. Cavanna, Y. Jin and G. Feve,  Nature Com.  {\bf  6}, 6854 (2015)

\bibitem{6Yuan} Y. L. Lim and A. Beige,  New. J. of Phys. {\bf 7}, 155 (2005)






\bibitem{17Kim}  H. Kim, O. Kwon and H. S. Moon, Sci. Reports. {\bf 9}, 18375 (2019)

\bibitem{18Kim}  H. Kim,  S. M. Lee and  H. S. Moon, Sci. Reports. {\bf 6}, 34805 (2016)

\bibitem{20Zhang}  Y. Z.  Zhang, K.  Wei, and F.  Xu. Phys. Rev. A, {bf 101}, 033823 (2020)

\bibitem{3Agne} S. Agne, J. Jin, K. B. Kuntz, F. M. Miatto, J. P. Bourgoin, and T. Jennewein  Opts. Exp. {\bf 28} 20943 (2020)


\bibitem{Campos1989} R. A. Campos, B. E. A. Saleh, and M. C. Teich, Phys. Rev. A  {\bf 40} 
, 1371 (1989)





\bibitem{12Feng} S. Feng and O. Pfister, Phys. Rev. Lett. {\bf 92}, 203601  (2004)
 



\bibitem{11Ferreri}  A. Ferreri, V. Ansa
ri, C. Silberhorn, and P. R. Sharapova, Phys. Rev. A,   {\bf 100} 053829 (2019)


\bibitem{16Kim}  Y. S.  Kim, O. Slattery, P.  S. Kuo, and X. Tang   Phys. Rev. A, {bf 87}, 063843 (2013)




\bibitem{13Kaltenbaek} R. Kaltenbaek, J. Lavoie, and K. J. Resch, Phys. Rev. Lett. {\bf 102}, 243601 (2009)


\bibitem{4Scott} H. Scott, D. Branford, N. Westerberg, J. Leach, and E. M. Gauger, Phys. Rev. A {\bf 102}, 033714 (2020)


\bibitem{5Lyons}  A. Lyons, G. C. Knee, T. Roger, J.  Leach , E .M. Gauger  and  D.  Faccio, Sci.  Adv. {\bf 4} 9416  (2018)



\end{references}
\end{document}